\documentclass{aa}  

\usepackage[version=4]{mhchem}
\usepackage[varg]{txfonts}
\usepackage{graphicx}
\usepackage{amssymb}                                
\usepackage[colorlinks,bookmarks]{hyperref}
\usepackage{color}
\usepackage[none]{hyphenat}
\definecolor{linkblue}{rgb}{0,0,0.8}
\definecolor{linkgreen}{rgb}{0,0.5,0}
\hypersetup{pdfpagemode=UseNone, pdfstartview=FitH, linkcolor=linkblue,citecolor=linkblue, urlcolor=linkblue}                    \usepackage{natbib}
\bibpunct{(}{)}{;}{a}{}{,} % to follow the A&A style            
\def\be{\begin{equation}} 
\def\ee{\end{equation}}

\def\gsim{\lower.5ex\hbox{\gtsima}} 
\def\lsim{\lower.5ex\hbox{\ltsima}} \def\gtsima{$\; \buildrel > \over 
\sim \;$} \def\ltsima{$\; \buildrel < \over \sim \;$} \def\prosima{$\; 
\buildrel \propto \over \sim \;$} \def\gsim{\lower.5ex\hbox{\gtsima}} 
\def\lsim{\lower.5ex\hbox{\ltsima}} 
\def\simgt{\lower.5ex\hbox{\gtsima}} 
\def\simlt{\lower.5ex\hbox{\ltsima}} 
\def\simpr{\lower.5ex\hbox{\prosima}}   
\def\gtsima{$\; \buildrel > \over \sim \;$} 
\def\ltsima{$\; \buildrel < \over \sim \;$} 
\def\gsim{\lower.5ex\hbox{\gtsima}} 
\def\lsim{\lower.5ex\hbox{\ltsima}} 
\def\simgt{\lower.5ex\hbox{\gtsima}} 
\def\simlt{\lower.5ex\hbox{\ltsima}} 
\def\simpr{\lower.5ex\hbox{\prosima}}

\def\E3{{\cal E}_{\rm g}^{III}}

\def\M*{M_*}

\def\Z*{Z_*}
\def\L*{L_*}

\defcitealias{Chatterjee_2026}{C26}

\makeatletter
\@ifundefined{linenumbers}{}{%
  \let\linenumbers\relax
  \let\nolinenumbers\relax
}
\makeatother

\begin{document}

%\title{Introduction of a primordial black hole in the SCRIPT framework}

\title{Impact of the primordial black hole population on 21-cm observables at high redshift}
%Assessment during cosmic dawn and reionisation
%%%%%%%%%%%%%%%%%%%%%%%%%%%%%%%%%%%%%%%%
% Please do not include ORCIDs next to author names.
% Only ORCIDs authenticated by individual authors in EDP Sciences editorial system will be taken into account.
% ORCIDs included here will be removed.
%%%%%%%%%%%%%%%%%%%%%%%%%%%%%%%%%%%%%%%%
\titlerunning{PBH in SCRIPT}
\author{   
Atrideb Chatterjee,\inst{1}\thanks{Corresponding Author.}
Barun Maity, \inst{2}\thanks{A. Chatterjee and B. Maity contributed equally to this work.}\and
Koushiki\inst{3}}
\institute{
Kapteyn Astronomical Institute, University of Groningen, PO Box 800, 9700 AV Groningen, The Netherlands \\ \email{a.chatterjee@rug.nl}
\and 
Max-Planck-Institut f\"{u}r Astronomie, K\"{o}nigstuhl 17, 69117 Heidelberg, Germany \\ \email{maity@mpia.de}
\and
International Centre for Space and Cosmology, School of Arts and Sciences, Ahmedabad University, Ahmedabad, GUJ 380009, India \\ \email{koushiki.malda@gmail.com}
}
% \abstract{}{}{}{}{}
% 5 {} token are mandatory
 
\abstract   
{The 21-centimetre (21-cm) signal is one of the most promising probes of the high-redshift Universe. It has typically been modelled without accounting for the effects of active galactic nuclei (AGNs) in the pre-JWST era, primarily due to the lack of observational evidence for AGNs at $z > 6$. However, following the discovery of several AGNs at redshifts as high as $z \sim 10$ by JWST, it has become imperative to incorporate the impact of these early AGNs when predicting the 21-cm signal. Based on the assumption that these AGNs are seeded by primordial black holes (PBHs), we studied their impact with a semi-numerical model setup. Specifically, we extended the explicitly photon-conserving reionisation framework, \texttt{SCRIPT}, including essential cosmic dawn physics and PBH contributions. This enabled us to compute both the global signal and the power spectrum of the 21-cm line over the redshift range 
$z \sim 30-5$ within a self-consistent framework. Building on this set-up, we investigated the impact of different PBH mass functions (obeying current observational constraints) on the resulting signal. We find the X-ray heating from PBHs can substantially make the global 21-cm signal shallower and suppress the expected power spectrum amplitude during cosmic dawn. We also find that the choice of mass function plays a crucial role in shaping the 21-cm signal, and can, in fact, lead to significantly different predictions.}


\maketitle
% **********************************************

%%%%%%%%%%%%%%%%% BODY OF PAPER %%%%%%%%%%%%%%%%%%

\section{Introduction}

 The 21-centimetre (21-cm) signal, originating from the hyperfine transition of neutral hydrogen, provides a powerful probe of the early Universe ($z \sim 30-5$) by tracing the thermal and ionisation state of the intergalactic medium (IGM) \citep[e.g.][]{Barkana_loeb_2001, furlanetto2006, pritchard2012, Mesinger_19}. Since the state of the IGM is governed by the properties of the astrophysical sources present at these epochs, the 21-cm signal is effectively determined by the properties of these early sources \citep[e.g.][]{Ghara_21, Chatterjee_21, Astraeus_I, Astraeus_VI}. 
 
Observationally, this signal is studied through two main aspects: $(i)$ the sky-averaged global signal and $(ii)$ its spatial fluctuations, commonly characterised by the power spectrum. A handful of studies in the past decade have considered accreting primordial black holes (PBHs) \citep[e.g.][]{Tashiro13, Gong_17, Mena_19, Yang_21}, Hawking-evaporating ultralight PBHs \citep{Mittal_22, Saha_22, Natwariya_2022}, astrophysical black holes (BHs) approximately tracing the star formation rate \citep{EwallWice_20}, and radiation fields scaling with the UV luminosity density \citep{Nelander2025}. However, both of these observational aspects have historically been modelled based on the assumption that the astrophysical sources are primarily star-forming (SF) in nature \citep[e.g.][]{Barkana_loeb_2001, furlanetto2006, Chatterjee_23, Dhandha_25, Sims_25, Gessey_25}, with potentially significant contribution from active galactic nuclei (AGNs) only at later times ($z \sim 5$) \citep{Onoue_2017, Dayal_20, Trebitsch_23, Dayal2025_Uncover}.
 
 In fact, the few studies mentioned above, which did include AGNs at early redshifts ($z>6$), were primarily aimed at explaining the non-standard feature of the global 21-cm signal tentatively detected by the EDGES experiment \citep{Bowman2018}, which has since been ruled out at $95.3\%$ confidence level by the Shaped Antenna Measurement of the Background Radio Spectrum 3 (SARAS-3) experiment \citep{Saras_22}. Recent James Webb Space Telescope (JWST) observations, revealing the presence of AGNs at epochs as early as $z \sim 6-10$, often dubbed as little red dots \citep{Schneider23, Greene_2024, bogdan2024, kovacs2024, napolitano2024, Napolitano_2025}, necessitates a reassessment of their potential impact on the 21-cm signal \citep{Kohri_22, Zhao_25, Zhao_26}. Assuming these early AGNs were seeded by PBHs \citep{liu2022, liu2023, gouttenoire2024, yuan2024, dayal2024_pbh, matteri2025, ziparo2025, Dayal_2026}, \citet[][\citetalias{Chatterjee_2026} hereafter]{Chatterjee_2026} developed a framework to incorporate the additional contribution of such PBH-seeded early AGNs in predicting the global 21-cm signal. We showed that these exotic objects can produce additional heating and make the absorption trough of the signal significantly shallower compared to the ones where only SF sources are included. Nevertheless, \citetalias{Chatterjee_2026} model had two key limitations:  $(i)$ it could only predict the globally averaged 21-cm signal but not its fluctuations (i.e, 21-cm power spectrum, due to its analytic nature) and $(ii)$ the global signal was computed under the assumption that PBHs follow a log-normal mass function.

 The aim of this work is to address both of these limitations. First, we extended the \citetalias{Chatterjee_2026} formalism to compute not only the global 21-cm signal but also the corresponding power spectrum. Second, we usde this extended framework to investigate the impact of different PBH mass functions on the 21-cm signal. As the PBH models producing these mass functions are different from each other in formation and evaporation mechanisms, each of these points towards vastly different primordial physical conditions. For instance, if the critical-collapse mass function is eventually found to be the dominant one, then we must assume that small pockets in the very early Universe have gone through a process of phase change.

With this in mind, we employed the Semi-numerical Code for ReIonization with PhoTon Conservation (\texttt{SCRIPT}), which incorporates a self-consistent treatment of inhomogeneous recombinations, the thermal evolution of the IGM, and radiative feedback effects during the epoch of reionisation (EoR) \citep{Maity_2022a}. We further extended this framework to include key cosmic dawn (CD) processes, such as X-ray heating and Ly-$\alpha$
 coupling, and incorporated the contribution from various PBH populations (inherently containing seeding mechanism information) on these physical quantities. This enables a unified treatment of IGM evolution across both the CD and the EoR, accounting for the effects of SF as well as PBH-seeded AGNs.

 This work is crucial given the ongoing and planned experiments to detect the 21-cm signal. While SARAS-3 \citep{Saras_22}, SCI-HI \citep{Voytek_2014}, the Broadband Instrument for Global Hydrogen Reionisation Signal (BIGHORNS; \citealt{Sokolowski_2015}), the Radio Experiment for the Analysis of Cosmic Hydrogen (REACH; \citealt{Reach_22}), and the Cosmic Twilight Polarimeter (CTP; \citealt{Nahn_2018}) are primarily focused on measuring the global signal, experiments such as the Low Frequency Array \citep[LOFAR;][]{LOFAR_13,Lofar_2025}, the Murchison Widefield Array \citep[MWA;][]{MWA_13,MWA_2025}, the Giant Metrewave Radio Telescope \citep[GMRT;][]{Paciga_2013}, the Hydrogen Epoch of Reionization Array \citep[HERA;][]{HERA_17,HERA_2023}, and the New Extension in Nançay Upgrading LOFAR \citep[NenuFAR;][]{NenuFAR_2024} are targeting the 21-cm power spectrum. Furthermore, the upcoming major facilities such as SKA-low (AA* and AA4 configuration) are expected to detect the fluctuation amplitude with percentage-level precision, along with providing a tomographic map of the 21-cm signal.

 Throughout this paper, we employ a $\Lambda$CDM model with dark
 energy, dark matter, and baryonic densities in units of the critical density as $\Omega_{\Lambda} = 0.691$, $\Omega_{m} = 0.308$, and $\Omega_{b} = 0.0482$, respectively, a Hubble constant of $H_{0} = 100 h \, \rm{km\, s^{-1}\, Mpc^{-1}}$ with $h=0.678$, spectral index of $n_{\rm s} = 0.961$ and normalisation of $\sigma_8 = 0.829$ \citep{Planck_2016}.

 The paper is organised as follows. Section\ref{sec:PBH_mf} describes different PBH mass functions and the semi-analytical model of the early AGNs seeded by these PBHs. Section-\ref{sec:sim_setup} describes the model and the simulation setup for SF galaxies. Section-\ref{sec:21cm_signal} presents the calculation of the 21-cm signal obtained from the contribution of both AGN and SF galaxies. Section-\ref{sec:results} describes the findings and, finally, we summarise our results and conclusion in Section-\ref{sec:results}.

%************
\section{PBH mass functions and early AGNs}
\label{sec:PBH_mf}

\subsection{PBH mass functions}

The discussion of the occurrence of BHs at a very early age of the Universe started with the works of Hawking and Carr \citep{Hawking:1971ei, Carr:1974nx}. They showed that BHs had the ability to form soon after the Big Bang $(\sim 10^{-47}s)$ and, thus, their production mechanism is very plausibly driven by inflationary potentials \citep{Mosani:2023vtr, Koushiki:2025dax} or phase transitions \citep{Carr:2020xqk}. Depending on whether the production mechanism follows inflationary or phase transitions, their mass function would be different \citep{Yokoyama:1998xd, Musco:2023dak}. In fact, even within an inflationary production mechanism, different inflationary potentials \citep{LINDE1983177, Lucchin1985, Albrecht1982} could lead to different PBH mass functions. 

The selection criteria for the sample are given below:

\begin{itemize}
    \item The formation of the PBHs, giving rise to the specific mass functions, must be triggered by a viable physical process. For instance, the mass function could arise from density fluctuations \citep{dolgov-silk1993}, or from phase-transitions \citep{Choptuik:1992jv, Koushiki:2025sce}. We note that there are multiple mass functions that are phenomenologically motivated \citep{ Chen:2016pud, Dienes:2025qdw}, but are not associated with a particular formation or collapse scenario. The latter category of mass functions is not physically motivated and therefore is not considered in this work. 
    
    \item These mass functions must also abide by observational constraints coming from spectral index distortion in CMB, low-scale lensing (micro, nano, femto) experiments, and accretion constraints \citep{Carr:2009jm}. The distributions, considered here, abide by these constraints \citep{Carr:2017jsz, Carr_21, carr-green2024}. 

    \item These mass functions must also remain high enough throughout the observational history for their abundance to be significant, even post-evaporation \cite{Carr:2020gox, Carr:2025kdk}. Evaporation is the mechanism by which the PBH loses mass gradually over time. This mechanism is quantum-gravitational in nature \citep{Goswami:2005fu, Filho:2025fvl} and a detailed discussion on this is beyond the scope of this paper. Without considering the effect of evaporation (which would be reflected through the loss of $M_{\rm PBH}$ over time) on how $\frac{dN}{d M_{\rm PBH}}$ changes dynamically, we used observational constraints coming from post-evaporation threshold of these mass functions, as they provide a natural way to segregate the physical ones from the physically unfeasible ones.
\end{itemize}

Among the set of plausible mass functions, only three distributions abide by all the constraints listed above. These PBH mass functions are briefly discussed below:

\begin{itemize}
    \item Log-normal mass function:
    In this case, PBHs are largely anticipated to originate from primordial fluctuations in space-time \citep{particles}. \cite{Dolgov:1992pu} proposed such a fluctuation of baryons producing higher densities at small scales and lower densities at larger scales. The small patches of such overdensities generate the fluctuations, which in turn produce PBHs with a mass function given by
    \be \label{eq: MSFlog} 
        \frac{dN}{d M_{\rm PBH}} = \frac{\kappa_{\rm LN}}{\sqrt{2\pi}\sigma M_{\rm PBH}^2 }\exp{\left(-\frac{{\ln{(M_{\rm PBH}/M_{\rm crit})}}^2}{2\sigma^2}\right)},
    \ee
    where $N$ is the number of PBH seeds in per unit co-moving volume, $M_{\rm PBH}$ is the mass of the PBH seed, $\kappa_{\rm LN}$ is the normalisation constant, and $\sigma$ is the deviation around the critical mass, $M_{\rm crit}$. We fixed $\sigma=0.7$ \citep{matteri2025} and $M_{\rm crit}= 10^{3.65} M_\odot$, which is the average mass of the PBH seed \citepalias{Chatterjee_2026}. 
    
    \item Power-law mass function:  \cite{Carr:1975qj} proposed another model of density fluctuation that changes as the effective equation-of-state parameter of the Universe changes. In this scenario, the fluctuation is added on top of an FLRW Universe. This specific form of fluctuation gives rise to the mass function,
    \be \label{eq: MSFpl}
        \frac{dN}{d M_{\rm PBH}} = \kappa_{\rm PL} M_{\rm PBH}^{-\alpha},
    \ee
    where $\kappa_{\rm PL}$ is the normalisation constant and $\alpha = \frac{2 (1+ 2\omega)}{(1+\omega)}$, with $\omega$ as the equation-of-state parameter of the Universe. In this work, we assumed that the PBH seeds are produced at the matter-dominated epoch, making $\omega =0$, which produces $\alpha = 2$.

    \item Critical mass function: A series of numerical and analytical studies has shown that for one-parameter families of scalar field initial conditions, gravitational collapse can lead to the formation of a zero-mass BH at a critical value of the parameter \citep{Choptuik:1992jv, Gundlach:1996eg, Koushiki:2025sc}. Later, this mechanism was used to explain the formation of PBHs via phase transitions \citep{Yokoyama:1998xd, Musco_2013} that produce mass function \citep{Carr:2017jsz}:
    \be \label{eq: MSFcrit}
        \frac{dN}{d M_{\rm PBH}} = \kappa_{\rm CL} M_{\rm PBH}^{1.85} \exp{\left[-\left(\frac{M_{\rm PBH}}{M_{H}}\right)^{2.85}\right]},
    \ee 
    where $\kappa_{\rm CL}$ is the normalisation constant and $M_{H}$ is the Hubble horizon mass \citep{Carr:2017jsz}, given by $M_{H}\equiv \frac{c^3 t_{\rm crit}}{G } = 2.03 \times 10^4 \left(\frac{t_{\rm crit}}{0.1 \, \rm s}\right) M_{\odot}$. Here, $t_{\rm crit}$ denotes the time of formation of PBH seeds due to critical collapse. Since PBHs may form as early as $10^{-47}\,\mathrm{s}$ after the Big Bang \citep{Carr_2003, Carr:2020gox, carr-green2024}, such seed formation is, in principle, possible at very early cosmological epochs. However, very early times lie beyond the regime of validity of classical dynamics owing to the expected importance of quantum gravitational effects. On the other hand, it would require $\gtrsim 1~\rm s$ to produce the supermassive BHs (SMBHs) that are typically found in galactic nuclei \citep{Carr_2003}. We therefore adopted $t_{\mathrm{crit}} = 0.1\,\mathrm{s}$ as a representative early-Universe epoch at which classical gravitational dynamics remains applicable, as well as allowing for sufficiently massive BH formation. We identified it with the time of formation of PBH seeds resulting from critical collapse. However, we note that the specific choice of $t_{\rm{crit}}$ does not affect the mass function (as long as it is reasonable) since its effect is absorbed in the normalising constant, $\kappa_{\rm CL}$.
    
    %\Kay{Phase transitions have been proposed as a possible mechanism underlying critical collapse \citep{Gundlach:1996eg,Chakrabarti:2025bsd}. The early Universe is expected to have undergone several phase transitions \citep{Garcia-Bellido:2021zgu}, each of which may potentially trigger critical collapse and the consequent formation of PBH seeds. Since PBHs may form as early as $10^{-47}\,\mathrm{s}$ after the Big Bang \citep{Carr:2020gox}, such seed formation is, in principle, possible at very early cosmological epochs. However, very early times lie beyond the regime of validity of classical dynamics owing to the expected importance of quantum gravitational effects. We therefore adopt $t_{\mathrm{crit}} = 0.1\,\mathrm{s}$ as a representative early-Universe epoch at which classical gravitational dynamics remains applicable, and identify it with the time of formation of PBH seeds resulting from critical collapse. It is to be noted that the specific choice of $t_{\mathrm{crit}}$ will not affect the mass function (as long as it is reasonable) and hereby, the results that we have obtained regarding 21cm signals.} \AC{Its effect will be absorbed in the normalising constant}
\end{itemize}
We note that all the mass functions have implicit dependence on time, since this mass grows as the PBH seed accretes mass with time as, discussed in detail in Section-\ref{sec:PHANES}.

We fixed the normalisation constants $\kappa_{\rm LN}, \; \kappa_{\rm PL}$, and $\kappa_{\rm CL}$, so  that the comoving number density of the PBHs matched the observed value of $10^{-5.27}$ $\rm cMpc^{-3}$ at $z\sim 10$ \citep{kovacs2024, bogdan2024} for a PBH seed with mass $10^{4.65} M_\odot$. The values for the normalisation constants come out to be $4.4 \times 10^{-8}$, $2.3 \times 10^{-1}$, and $4.2\times 10^{-15}$ for log-normal, power-law, and critical mass functions, respectively.

\subsection{Semi-analytical Model for early AGNs}
\label{sec:PHANES}

To describe the formation and evolution of the early AGNs seeded by these PBHs, we followed the \texttt{PHANES} framework closely \citep{Dayal_2026}. We note that it was later extended by \citetalias{Chatterjee_2026} to derive different quantities (e.g. X-ray emissivity, Ly-$\alpha$ production rate, and ionising photon production rate) that are critical for obtaining the 21-cm signal. Following \citetalias{Chatterjee_2026}, the effect of astrophysically produced AGNs, is ignored in this work. It follows from the fact that these AGNs start to affect the 21-cm signal around $z \sim 5$ \citep{Dayal2025_Uncover}, whereas the amplitude of the 21-cm signal (both global and power spectra) approaches zero by this redshift.

Here, we briefly describe the important features of this framework, emphasising the calculation of key quantities related to the 21-cm signal.

\begin{enumerate}

\item In this model, the PBH seeds in the mass-range $10^{0.6} - 10^{6}$ $M_{\odot}$ \citep{Dayal_2026}, forming at $z \sim 3400$, initially grow by linear accretion of dark matter around themselves until $z \sim 34$. Afterwards, the halo growth becomes non-linear and finally begins to accrete gas once its baryonic overdensity reaches $200$, triggering the star formation. At this point, the model accounts for feedback from both BH accretion and star formation to calculate the subsequent evolution of the system until $z \sim 5$.

\item The bolometric luminosity of such an individual AGN at redshift, z, is computed as

\be \label{lbol}
L_{\rm bol} = \epsilon_{r} \frac{\Delta M_{\rm PBH}~ c^2}{\Delta t},
\ee 
where $\Delta M_{\rm PBH}$ is the mass accreted by the PBH seed after each time step of $\Delta t = 20 \, \rm Myr$ \citep{dayal2024_pbh}. Assuming the BHs to be non-spinning, the radiative efficiency, $\epsilon_{r}$, is fixed at $0.057$ to match existing astrophysical and cosmological constraints. \footnote{Note that, the remaining free parameters in the \texttt{PHANES} framework related to AGN and stellar feedback are adopted from Table-1 of \cite{Dayal_2026}, corresponding to the non-spinning BH scenario.}

\item From a given $L_{\rm bol}$, the X-ray luminosity, $L_{X}$, can be obtained using the bolometric correction factor $k_{X}$, defined as $k_{X}= L_{\rm bol}/L_{X}$. Following \cite{duras2020}, $k_X$ is computed from the fitting function $k_{X} = a\left[ 1 + \left(\frac{\log_{10}(L_{\rm bol}/L_{\odot}) }{b}\right)^c \right]$, with the best fit values of the free parameters given by $a=10.96$, $b=11.93$, $c=17.79$. Once we calculate the $L_{\rm X}$ values of the individual AGNs, the global X-ray emissivity is obtained using
\begin{equation}
\label{eq: eX_PBH}
\langle \epsilon^{\rm PBH}_{X} \rangle = \int \frac{dN}{d M_{\rm PBH}} \cdot L^{\rm PBH}_{X}\, dM_{\rm PBH}\;.
\end{equation}

\item The globally averaged Ly$-\alpha$ photon production rate by these AGNs at redshift, z, is given by
\be \label{nuphot}
\langle \dot{n}^{\rm PBH}_{\alpha} \rangle = \int \frac{dN}{dM_{\rm PBH}} \frac{L_\alpha}{h_p \nu_{\rm \alpha}} dM_{\rm PBH},
\ee 
where $h_{p}$ is the Planck constant, $\nu_{\alpha}$ is the frequency corresponding to Ly$-\alpha$ photons, and $L_\alpha$ is the Ly$-\alpha$ luminosity of each AGN obtained from the B-band luminosity, $L_{B}$ assuming a power law index of $-0.57$ \citep{Lusso_15}. Further, $L_{B}$ is calculated from $L_{\rm bol}$ using the fitting function from \cite{marconi2004}. A similar approach has been taken for calculating the number of ionising photons from these PBH-seeded AGNs except with a power law index of $-1.57$ \citep{Telfer_02}. As shown later in this work, the ionising contribution from these early AGNs is negligible compared to the SF galaxies.
\end{enumerate}

\begin{figure}
    \includegraphics[width=\linewidth]{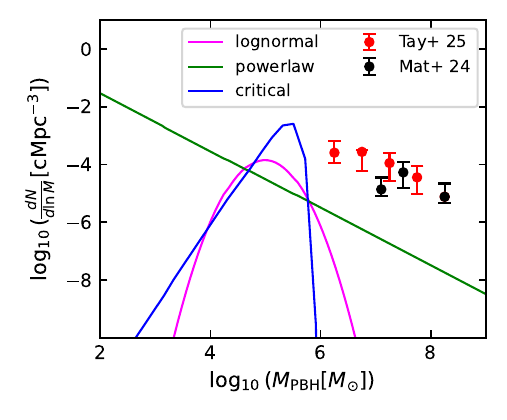}
    \caption{Mass functions obtained for different PBH models at $z \sim 5$ along with the recent JWST estimates \citep{matthee2024, taylor2025}. The magenta, green, and blue lines correspond to power-law, log-normal, and critical mass functions, respectively. Furthermore, the black points with error bars are from \cite{matthee2024}, whereas the red ones are taken from \cite{taylor2025}. Since we do not add the contribution from astrophysically produced AGNs here, all the PBH models produce mass functions that are two to three orders of magnitude lower than the observed values.}
    \label{fig:BH_massfunction}
\end{figure}

In Figure-\ref{fig:BH_massfunction}, we show the mass functions obtained for three different PBH models, at $z\sim 5$ using the aforementioned analytical calculation. We start by noting that the log-normal (in magenta) and critical (in blue) mass functions show their peaks around $M_{\rm BH} \sim 10^5 M_{\odot}$, whereas their values are negligible in both higher  $( M_{\rm BH} > 10^7 M_{\odot} )$ and lower $(M_{\rm BH} < 10^3 M_{\odot})$ mass regime. This is expected from their functional forms given in equations \ref{eq: MSFlog} and \ref{eq: MSFcrit}. We show that although the power-law (in green) mass function is dominated by the contribution from the lower mass BHs, it has a non-negligible contribution even from higher mass bins. As a sanity check, we show that each of these three mass functions is 2-3 orders of magnitude lower than that of the mass functions observed by JWST \citep{matthee2024, taylor2025}. This is only to be expected since we do not consider the contribution coming from astrophysically produced AGNs while computing the mass functions.

% ********************   
\section{Simulation setup with SCRIPT} \label{sec:sim_setup}
Next, we briefly describe the simulation setup employed in this study. We utilised an explicitly photon-conserving semi-numerical reionisation model, \texttt{SCRIPT}, and extended its framework to incorporate the essential physics of the cosmic dawn. The baseline model was introduced in \citet{choudhury_paranjape22} and subsequently enhanced with key reionisation processes, including inhomogeneous recombination and radiative feedback, as presented in \citet{Maity_2022a}. 

The model self-consistently predicts the thermal and ionisation state of the Universe within a cosmologically representative simulation volume. It takes as input the density field ($\Delta_i$, where $i$ labels individual cells in the simulation grid) and the collapsed mass fraction ($f_{\mathrm{coll},i}$). These quantities were used to compute the effective number of ionising photons through an astrophysical parametrisation of the ionising efficiency, $\zeta$. The product $\zeta f_{\mathrm{coll},i}$ represents the number of ionising photons per baryon in each cell. These photons are then distributed across the source cell and its neighbouring cells to determine the ionisation fraction, while explicitly enforcing photon conservation.

This effective photon budget is updated self-consistently in each cell to account for inhomogeneous recombination and radiative feedback. The recombination rate is computed by tracking the joint evolution of the ionisation and density fields with redshift. To incorporate the impact of small-scale clustering of photon sinks, we adopt a globally averaged clumping factor ($C_{\mathrm{HII}}=3$), consistent with results from hydrodynamical simulations \citep[e.g.][]{DAloisio_2020}. The recombination rates further depend on the thermal history of the gas, as discussed in Section \ref{sec:21cm_signal}. Photoionisation heating can be modelled by parametrising the temperature increase ($T_{\mathrm{re}}$) when a region transitions from neutral to ionised state \citep{Hui_1997, Furlanetto_Oh_2009, Keating_2018, Maity_2022a, Maity_choudhury_2022b}.

Radiative feedback is incorporated by modifying the minimum halo mass threshold ($M_{\mathrm{min},i}$) for star formation. In neutral regions, this threshold is set by the atomic cooling limit. In contrast, in ionised regions it is determined by the Jeans mass at the virial overdensity ($M_{J,i}$). As the Jeans mass depends on the local temperature ($M_{J,i} \propto T_i^{3/2}$), our self-consistent treatment of thermal evolution allows us to dynamically update $M_{\mathrm{min},i}$ (and, consequently, $f_{\mathrm{coll},i}$), thereby capturing the effects of radiative feedback.

As our focus is on large-scale observables, we generate the density field using second-order Lagrangian perturbation theory (2LPT) \citep[\texttt{MUSIC},][]{music_2011}, rather than computationally expensive $N$-body simulations.. The collapsed mass fraction is then estimated using a semi-analytic model based on conditional ellipsoidal collapse \citep{sheth-tormen2002}, which has been detailed in \citet{choudhury_paranjape22}. For this work, we adopted a simulation volume of a size of $L = 256~h^{-1}\mathrm{cMpc}$ with a spatial resolution of $\Delta x = 8~h^{-1}\mathrm{cMpc}$. This box size is sufficient to capture the scales relevant for 21-cm observables \citep{Iliev_2014, Kaur_2020}.
To model the evolution of the IGM during cosmic dawn and reionisation, we generated 251 coeval simulation boxes spanning the redshift range $z = 5$–30, with a resolution of $\Delta z = 0.1$.

Below, we discuss the essential model components and their astrophysical significance. This further provides a physical background for the inclusion of the PBH contribution in the existing model set-up.

\begin{enumerate}
\item The number of escaping ionising photons produced per unit time at redshift, $z$, is computed using
\begin{equation}
\begin{split}
       \dot{n}_{\mathrm{ion},i}  &= \frac{dN_{\rm ion}}{dM} \times f_{\mathrm{esc}} \times\rho_{\mathrm{SFR}, i},\\
       &= \bar{\rho}_{\rm b} \left( \frac{dN_{\rm ion}}{dM} \right) \frac{d}{dt}\langle f_{*}f_{\mathrm{esc}} f_{{\rm coll}, i} \Delta_{i}\rangle,
\end{split}
\end{equation}
where $f_{*}$ is the star formation efficiency, $f_{\rm esc}$ is the escape fraction, and $\bar{\rho}_{\rm b}$ is the mean comoving baryon density. Finally, $\frac{dN_{\rm ion}}{dM}$ is the number of ionising photons produced per unit stellar mass. Following the formalism developed in \cite{Maity_2022a}, and defining $\zeta = 1.22 f_{*}f_{\rm esc}N_{\gamma}$, we can rewrite the above equation as
\begin{equation}
\dot{n}_{\mathrm{ion},i} =\frac{\bar{n}_{\rm b}}{1.22}\frac{ d\left\langle\zeta  f_{\mathrm{coll},i} \Delta_i  \right\rangle}{d t} =  \bar{n}_{H}\frac{ d\left\langle\zeta  f_{\mathrm{coll},i} \Delta_i  \right\rangle}{d t},
\end{equation}

where $N_{\gamma}$ is the number of ionising photons per unit baryon, $\bar{n}_{b}$ ($\bar{n}_{H}$) is the average baryon (hydrogen) number density. We adopted $N_{\gamma}=4845$, motivated by typical values from \texttt{Starburst99} \citep{Starburst_99} assuming a standard Salpeter IMF in the mass range $1 - 100$ $M_{\odot}$ with a metallicity of $0.05 M_{\odot}$. We further assumed a redshift-dependent star formation efficiency, $f_*(z)=f_{*,0}~\left(\frac{10}{1+z}\right)^{\alpha_z}$ \citep{Sun_16,Mirocha_2017, Choudhury_21, Maity_2022a}. For the fiducial model, we assumed $f_{*,0}=0.005$, $\alpha_z=4$, and fixed the value of the ionising escape fraction, $f_{\rm esc} = 0.12$ \citep{Maity_choudhury_2022b}, ensuring that the model matches the existing constraints on reionisation history (reionisation start at $z\sim10$ and ends at $z\sim5.7$) as shown in Appendix-\ref{fig:ion_hist}. As we aim to study the effect of PBH during cosmic dawn,  a relatively sharp reionisation history has been assumed to keep the cosmic dawn unaffected by the photoionisation heating.

\item Following a similar definition, the Ly-$\alpha$ photon production rate at redshift, $z$, in the $i^{th}$ cell is given by
\begin{equation}
 \dot{n}_{\alpha,i}^{\rm SF} = \bar{\rho}_{b} \left( \frac{dN_{\rm \alpha}}{dM} \right)\frac{ d\left\langle f_{*}f_{\mathrm{esc}} f_{{\rm coll}, i} \Delta_{i}  \right\rangle}{dt}
 =\bar{n}_{b}N_{\rm \gamma,\alpha}\frac{ d\left\langle f_{*}f_{\mathrm{esc}} f_{{\rm coll}, i} \Delta_{i}  \right\rangle}{dt}\label{eq:Lyalpha_SF},
\end{equation}
where $N_{\rm \gamma,\alpha}$ is the number of Ly-$\alpha$ photons per unit baryon, which is assumed to be twice that of ionising photons (i.e. 9690). We further added 4800, taking into account the Pop-III stellar contribution  \citep{Barkana_05, furlanetto2006}.

\item The evolution of the global ionisation fraction ($\rm Q_{\rm HII}$) can be written as
\begin{equation}
\begin{split}
    \frac{d Q_{\mathrm{HII}, i}}{dt} = \frac{ d\left\langle\zeta  f_{\mathrm{coll},i} \Delta_i  \right\rangle}{d t} - Q_{\mathrm{HII}, i}\alpha_{A} \, C_{\rm HII}\, \chi_{\rm He}\bar{n}_{H} (1+z)^3,
\end{split}
\end{equation}
where $\chi_{He}$ is the correction factor to electron number density due to helium ionisation, $C_{\rm HII}$ is the clumping factor of the IGM (fixed at 3), and $\alpha_{A}$ is the (case A) recombination rate coefficient. We note that we do not consider the contribution of PBHs while calculating the ionisation history as the ionising photons coming from different PBH models are entirely negligible compared to those of SF galaxies (discussed in Section-\ref{sec:results}).
\end{enumerate}

% ***********************

\section{The 21-cm signal}
\label{sec:21cm_signal}

The 21-cm differential brightness temperature 
\citep{Pritchard_2012, Mena_19} can be written as
\begin{equation}
      \delta T_{b,i}   \simeq  27 \, {\rm mK}\,\, \,  x_{\mathrm{HI}, i}\Delta_{i} \left( \frac{\Omega_bh^2}{0.023}\right)\left(\frac{1+z}{10} \frac{0.15}{\Omega_mh^2}\right)^{1/2}\left( 1- \frac{T_{\gamma, i}}{T_{S, i}} \right) 
\end{equation}
where $x_{\mathrm{HI}, i}$ denotes the neutral hydrogen fraction of the IGM and $T_{\gamma, i}(z)$ is the background CMB temperature given by $T_{\gamma, i} \equiv T_{\gamma} = 2.73\,(1+z)\, {\rm K}$. Finally, $T_{S, i}$, the spin temperature of the hydrogen atom, which can be written as
\begin{equation}  
T_{S,i}^{-1}=\frac{T_{\gamma}^{-1}+x_{\alpha, i}T^{-1}_{K, i} + x_{c, i}T^{-1}_{K,i}}{1+ x_{c, i} + x_{\alpha, i}},
    \label{eq:tspin}
\end{equation}
where $x_{\alpha}$ is the Ly-$\alpha$ coupling coefficient, $x_{c}$ is the collisional coupling coefficient, and $T_{K, i}$ is the kinetic temperature of the IGM.

Another observable of our interest for this study is the dimensionless 21-cm power spectrum, defined as
\begin{equation}
\Delta_{21}^2 = \frac{k^3P_{21}(k)}{2 \pi^2},
\end{equation}
where $P_{21}(k)$ is the power spectrum of the mean-subtracted
fluctuation field $\delta T_{b,i}-\langle \delta T_{b,i} \rangle$.

The collisional coupling coefficient, $x_{c,i}$ is determined using the standard fitting functions described in \cite{pritchard2012}, whereas all the other terms appearing in equation \ref{eq:tspin} are described below.

\subsection{Kinetic temperature of the IGM}
The temperature evolution of the IGM can be expressed as 

\begin{equation}
    \frac{dT_{K,i}}{dz} = \frac{2 T_{K,i}}{1+z}+ \frac{dt}{dz}\frac{2}{3} \frac{\epsilon_{i}}{k_Bn_{\mathrm{tot},i}},
\end{equation}
where the first term on the right-hand side denotes the adiabatic cooling due to the expansion of the Universe, and $\epsilon_{i}$ \textbf{in} the second term represents the net heating rate per unit volume for the $i^{\rm th}$ cell from different astrophysical processes. While we did follow \cite{furlanetto2006} for computing the adiabatic cooling, here we explain in detail the calculation of the second term, which can be expanded 
\citep{furlanetto2006, Maity_2022a} via
\begin{equation}
\begin{split}
    \frac{2}{3}\frac{\epsilon_{i}}{k_Bn_{\mathrm{tot},i}} & = \frac{2}{3}\frac{1}{k_Bn_{\mathrm{tot}, i}}\left( \epsilon^{\rm SF+PBH}_{X,i} + \epsilon^{\rm SF}_{\mathrm{re}, i} + \epsilon_{\mathrm{comp}, i}\right),
\end{split}    
\end{equation}
where the first term denotes the total X-ray heating of the IGM coming due to both SF and PBH-seeded galaxies. Furthermore, $\epsilon_{\mathrm{re}, i}$ represents the heating due to reionisation, and finally $\epsilon_{\mathrm{comp}, i}$ indicate Compton heating and cooling. We followed \cite{Maity_2022a} in order to determine $\epsilon_{\mathrm{re}, i}$, and $\epsilon_{\mathrm{comp}, i}$. Specifically, $\epsilon_{\mathrm{re}, i}$ is essentially proportional to the reionisation temperature increment ($T_{\rm re}$) parameter (mentioned in Section-\ref{sec:sim_setup}). We note that the formalism also includes Ly$-\alpha$ heating contribution \citep{furlanetto2006, Mittal_21}; however, that is negligible in comparison to X-ray heating and does not affect any of the  quantities significantly in our assumed fiducial model. Finally, combining the contribution of the SF and PBH galaxies, the total X-ray heating term (i.e. $\epsilon^{\rm SF+PBH}_{X,i}$;  \citetalias{Chatterjee_2026}) can be written as

\begin{equation}
    \epsilon^{\rm SF+PBH}_{X,i} = f_{h}\left[f^{\rm SF}_{X, \rm esc}\epsilon^{\rm SF}_{X} + f^{\rm PBH}_{X, \rm esc}\epsilon^{\rm PBH}_{X}\right],
\end{equation}
where $f_{h}$ is the fraction of X-ray that heats the IGM and is taken to be $\sim \left(1+\frac{2x_{\mathrm{HII},i}}{3}\right)$, with $x_{\mathrm{HII},i}$ being the ionisation fraction \citep{furlanetto2006}. Furthermore, $f^{\rm SF}_{X, \rm esc}$ and $f^{\rm PBH}_{X, \rm esc}$ represent the fraction of X-rays escaping the host galaxies for SF and PBH seeded systems. \footnote{We also note that these factors will absorb any uncertainty that may come from using the low-redshift $\epsilon_X - \rho_{\rm SFR}$ calibration relation while determining the X-ray contribution from SF and PBH galaxies.}. As mentioned later in this work, we assume $f^{\rm SF}_{X, \rm esc}$ and $f^{\rm PBH}_{X, \rm esc}$ to be $1.0$ (\citetalias{Chatterjee_2026}) for the fiducial model. Next, we describe each component of this term.

X-ray from SF: Assuming that the relation between X-ray luminosity and the star formation rate density ($\rho_{\rm SFR}$) in the Local Universe holds also in the high-z Universe, the X-ray emissivity \citep{mineo2012, furlanetto2006} can be expressed as

\iffalse \begin{equation}
\begin{split}
    \frac{\epsilon^{\rm SF}_{X,i}}{{\rm erg \, sec^{-1}\, \,Mpc^{-3}}} &=  3.4 \times 10^{40} \times  \rho_{{\rm SFR}, i}, \;
\end{split}
\end{equation}
\fi
\begin{equation}
\begin{split}
    \frac{2\epsilon^{\rm SF}_{X,i}}{3k_Bn_{\mathrm{tot},i}H(z)} &=  5\times 10^3~ {\rm K} \left(\frac{f_*}{0.1}\frac{\mathrm{d} f_{\rm coll}/\mathrm{d} z}{0.01}\frac{1+z}.{10}\right)\;
\end{split}
\end{equation}

X-rays from PBH: For PBH, the X-ray emissivity in the $i^{\rm th}$ cell is given by
\begin{equation}
    \epsilon^{\rm PBH}_{X,i} = \Delta_i \langle \epsilon^{\rm PBH}_{X} \rangle,
\end{equation}
where we assume that the X-ray emissivity in the $i^{\rm th}$ cell can be obtained by multiplying the global X-ray emissivity $\langle \epsilon^{\rm PBH}_{X} \rangle$, from Equation \ref{eq: eX_PBH}, by the overdensity of the $i^{\rm th}$ cell. Given that the formation of PBHs is related to the collapse of overdense regions in the primordial Universe, this equation provides a physically motivated form introducing spatial variation in X-ray emissivity of PBHs.

\subsection{Ly-$\alpha$ coupling}
The Ly-$\alpha$ coupling coefficient $x_{\alpha, i}$ is defined as
\begin{equation}
    x_{\alpha, i} = 1.81 \times 10^{11} \, (1+z)^{-1} S_{\alpha} \frac{J_{\alpha, i}}{\rm cm^{-2} sec^{-1} Hz^{-1} sr^{-1}},
\end{equation}
where $J_{\alpha,i}$ is the Ly-$\alpha$ background flux, $S_{\alpha}$ is a factor coming from a detailed analysis of atomic physics and, following the general practice, we take it to be unity ignoring the effect of spectral distortions \citep{furlanetto2006} (also see \citep{Hirata_06, Chuzhoy_06, Mittal_21} where the authors consider a more detailed calculation to determine the value of $S_{\alpha}$ and show that the value of $S_{\alpha}$ may deviate from $1.0$).

Finally, including the effect of both SF and PBH-seeded galaxies, the background Ly-$\alpha$ flux  \citep{Mena_19} can be written as

\begin{equation}
    J_{\alpha, i}^{\rm SF+PBH}=\frac{c}{4 \pi}\frac{1}{H(z)\nu_{\alpha}} \left[f^{\rm SF}_{\alpha}\dot{n}^{\rm SF}_{\alpha, i}(z)+ f^{\rm PBH}_{\alpha}\dot{n}^{\rm PBH}_{\alpha, i}(z)\right],
    \label{eq:Ly_alpha background}
\end{equation}
with $\dot{n}^{\rm SF}_{\alpha, i}(z)$ given by equation \ref{eq:Lyalpha_SF} and $\dot{n}^{\rm PBH}_{\alpha, i} =  \Delta_{i}\langle n^{\rm PBH}_{\alpha} \rangle$. Then, $f^{\rm SF}_{\alpha}$ and $f^{\rm PBH}_{\alpha}$ are the escaping fraction of the Ly-$\alpha$ photons from SF and PBH-seeded galaxies. For the sake of simplicity, we assumed both of them to be 1.

\iffalse \subsection{reionisation}

The neutral hydrogen fraction, $x_{\rm HI}$, determined from the evolution of the volume filling factor for ionised hydrogen ($\rm Q_{\rm HII}$) can be written as
\begin{equation}
\begin{split}
    \frac{d Q_{\rm HII, i}}{dt} = \frac{f_{\rm esc}\dot{n}^{\rm SF}_{\rm ion, i}}{n_{\rm H, com}} - Q_{\rm HII, i}\alpha_{B} \, {\cal C}\, n_{\rm H, com} (1+z)^3,
\end{split}
\end{equation}
where $ n_{\rm H, com}$ is the hydrogen comoving number density,  ${\cal C}$ is the clumping factor of the IGM, and $\alpha_{B}$ is the (case B) recombination rate coefficient. The functional form of clumping factor C is taken as $1+43z^{-1.71}$ \citep{pawlik2009}. Finally, $\dot{n}^{\rm SF}_{\rm ion}$ and $\dot{n}^{\rm PBH}_{\rm ion, esc}$ denote the rate of ionising photons produced in SF and PBH-seeded galaxies, respectively. As a sanity check of whether the extended framework of \texttt{SCRIPT}  has been implemented properly, we have shown the 21-cm signal (global as well as the power spectrum) with varying $f_{*,0}$ in Appendix-\ref{app:reion_hist}. We note that we did not consider the contribution of PBH galaxies while calculating the ionisation history, as the ionising photons coming from different PBH models are completely negligible compared to SF galaxies (discussed later in Section-\ref{sec:results}). \fi

%\Kay{The spacings between paragraphs have to be made uniform.}

\section{Effect of PBHs on 21-cm signal}
\label{sec:results}
\begin{figure*}
    \includegraphics[width =\linewidth]{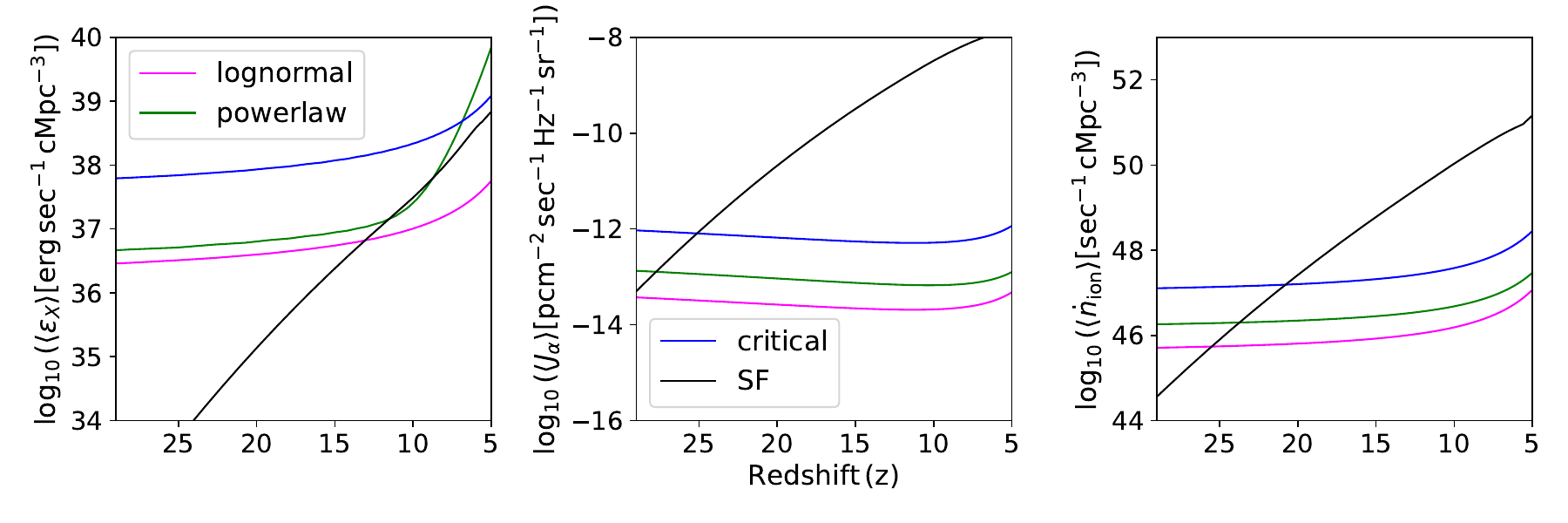}
    \caption{Redshift evolution of the global X-ray emissivity, Ly-\(\alpha\) background flux (in units of physical (p) $\rm cm^{-2} ~ sec^{-1} ~ Hz^{-1} ~ sr^{-1}$) and ionising photon production rate for different PBH models along with the SF galaxies. The black, magenta, green, and blue lines, respectively, denote the SF, lognormal, power-law, and critical PBH seeded systems as noted in the legends.}
    \label{fig:all_key_quantities}
\end{figure*}

We begin this section by describing Figure-\ref{fig:all_key_quantities}, which illustrates the redshift evolution of the quantities critical for the determination of the 21-cm signal. In all panels, the magenta, green, and blue curves correspond to the lognormal, power-law, and critical PBH mass functions, respectively, while the black curve represents the contribution from SF galaxies (for a fiducial scenario).

In the left-most panel, we show the redshift evolution of the X-ray emissivity for different models studied in this work. First, we note that all PBH models produce significantly higher X-ray emission than SF galaxies at $z \geq 15$, reflecting the earlier onset of X-ray production in PBH-seeded systems compared to SF galaxies. In the range $10 \leq z < 15$, the critical mass function produces the highest X-ray emissivity, followed by the power-law model and SF galaxies, while the lognormal PBH model yields the lowest emission. Keeping in mind that the emission from individual galaxies is independent of the specific mass function, this behaviour can be qualitatively understood from the overall shapes of the mass functions of different PBH models shown in Figure-\ref{fig:BH_massfunction}. The critical mass function, with its highest peak, yields the strongest emission. Although the lognormal mass function attains a higher peak than the power-law form, the latter’s extended tail results in a greater overall contribution, ultimately surpassing the lognormal mass function. Finally, at $z \lesssim 8$, the X-ray emission from power-law PBHs exceeds that of the critical mass function. This is attributed to the extended high-mass tail of the power-law distribution, whereas the critical mass function is sharply peaked and contributes negligibly at higher masses.    

The centre and right-most panels show the redshift evolution of the background Ly-$\alpha$ flux ($J_{\alpha}$ in units of physical $\rm cm^{-2} sec^{-1} Hz^{-1} \, sr^{-1}$) and the ionising photon production rate $(\dot{n}_{\rm ion})$, respectively. Since both Ly-$\alpha$ and ionising photons are produced predominantly via star formation, it is expected that SF galaxies generate Ly-$\alpha$ and ionising photons several orders of magnitude higher than any of the PBH models over the entire redshift range. Furthermore, among different PBH models, the critical mass function yields the most, followed by the power-law mass function, whereas the lognormal PBH model produces the lowest emission. Similar to the explanation of the X-ray emission in the left-most panel, this can again be attributed to the shape and normalisation of different mass functions, as shown in Figure-\ref{fig:BH_massfunction}.

\begin{figure*}
\sidecaption
    \includegraphics[width=12cm]{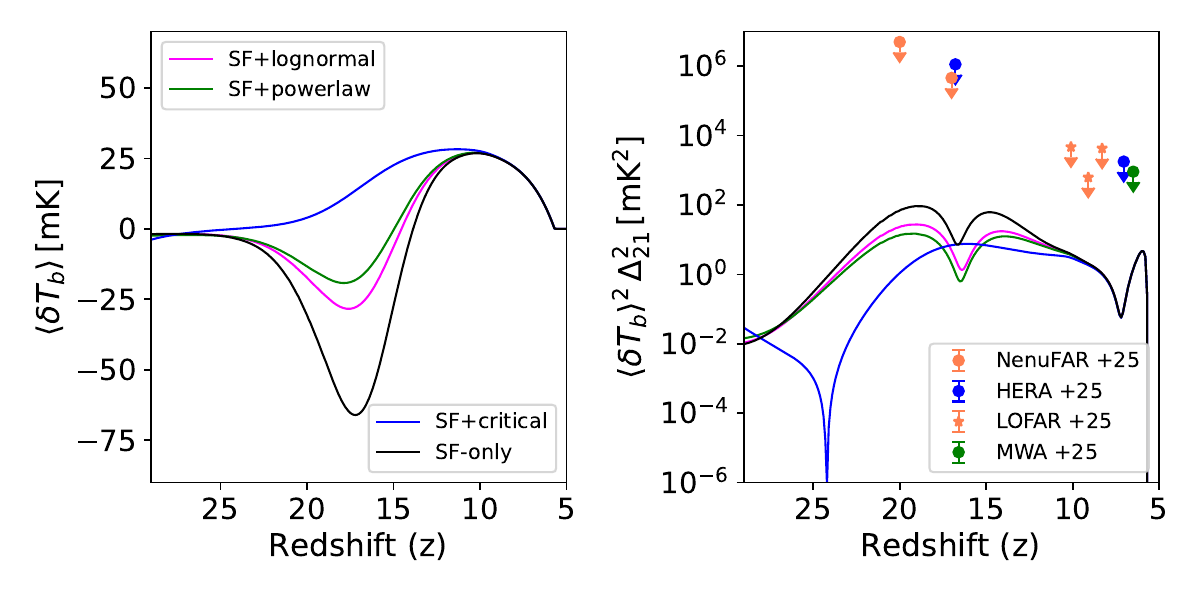}
    \caption{Redshift evolution of the 21-cm global signal and the power spectrum (computed at $k \sim 0.1 \,h\,\rm cMpc^{-1})$ corresponding to different models studied in this paper. The legends in the left panel show the colour scheme for various models, whereas the legends in the right panel show the upper limits obtained from different observations.}
    \label{fig:global_PS}
\end{figure*}

Now, we move on to our main results, depicted in Figure-\ref{fig:global_PS}. The left panel shows the redshift evolution of the global 21-cm signal corresponding to different models, whereas the right panel shows the observable for the 21-cm power spectrum, $\langle \delta T_{b}\rangle^2\Delta^2_{21}$ computed at $k \sim 0.1 \,h\,\rm cMpc^{-1}$ in the entire redshift range of our interest. Before going into the detailed discussion, we recall that for all these models depicted in this figure, we have $f^{\rm SF}_{\alpha} = f^{\rm PBH}_{\alpha} = f^{\rm SF}_{X, \rm esc} = f^{\rm PBH}_{X, \rm esc} = 1.0$. 

For the global signal, the lognormal (in magenta) and power-law PBHs (in green) produce similar redshift evolution with absorption trough of $\sim -30 ~mK$ at a redshift of $z \sim 18$, whereas the 21-cm global signal due to PBHs following critical mass functions (in blue) is very distinct compared to both of these models. In fact, it does not even produce any absorption trough. This behaviour can be explained from a closer inspection of the X-ray emissivity produced from different models shown in the left-most panel of Figure-\ref{fig:all_key_quantities}. In Figure-\ref{fig:all_key_quantities}, we find that the X-ray emissivity is highest for critical and lowest for the lognormal, with a power-law model between them. Keeping in mind that the depth of the signal is effectively determined by the X-ray heating in the IGM for these systems (as the Ly-$\alpha$ background is very similar for all the PBH models), the model with the maximum X-ray heating (i.e. critical mass function) will produce the shallowest trough, while the model with the lowest X-ray emission (i.e. lognormal mass function) will produce the deepest trough. Interestingly, we also note a slightly higher negative value of global signal (and a higher power spectrum amplitude) for the critical mass function in comparison to the other two, even at $z \sim 30$ due to a very high Ly-$\alpha$ background resulting in early onset of Ly-$\alpha$ coupling. Finally, we also show the 21-cm signal produced from an SF-only model (in black) with no contribution from PBH-seeded systems. As expected, the SF-only model shows the deepest absorption trough of $\sim -60$ mK without any X-ray heating coming from PBH-seeded systems. The finding that models with PBH contribution will show a shallower absorption trough compared to models without PBH contribution has also been found in \citetalias{Chatterjee_2026}.

The power spectra (the right panel of Figure-\ref{fig:global_PS}) contain three distinct peaks corresponding to Ly-$\alpha$ coupling, X-ray heating, and reionisation heating, respectively, at higher to lower redshifts. It is apparent that the PBH heating suppresses the amplitude of the power spectra during the cosmic dawn. Specifically, the power law and lognormal cases show more than a factor of half reduction on the power spectrum amplitude ($\sim20-30~mK^2$ from the SF-only case $\sim100~mK^2$). This effect is more striking for the critical case, where a suppression of more than a couple of orders of magnitude can be seen. However, all the scenarios are still allowed by the existing limits from the recent interferometric observations. The qualitative effects are consistent with the previous studies \citep{Mena_19}. We note that PBHs do not affect the power-spectrum amplitude during the reionisation epoch, as they do not contribute to ionisation in our model. However, these effects can be highly degenerate with the photoionisation heating if the reionisation is extended to cosmic dawn (unlike our fiducial case).

\begin{figure*}
\sidecaption
    \includegraphics[width=12cm]{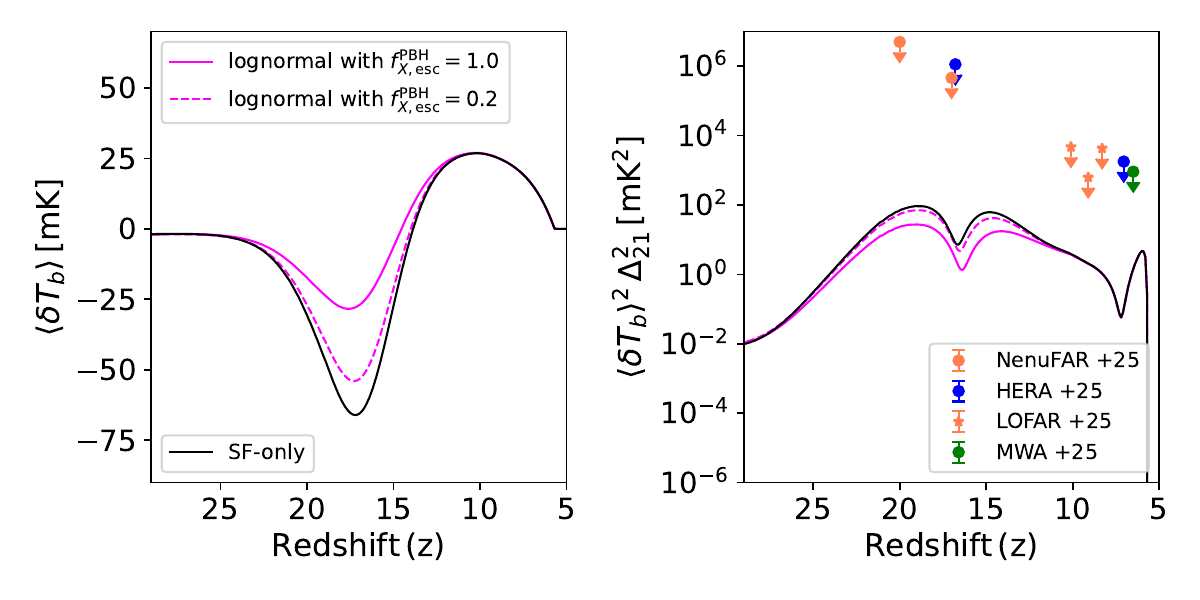}
    \caption{Effect of varying $f^{\rm PBH}_{X,\rm esc}$ in the lognormal mass function. As mentioned in the legend, the dashed magenta line shows the scenario with $f^{\rm PBH}_{X, \rm esc} = 1.0$, whereas the solid magenta line represents the case with $f^{\rm PBH}_{X, \rm esc} =1.0$. Further, the black line shows the SF-Only scenario.}
    \label{fig:global_ps_f_X_variation}
\end{figure*}

Another important aspect of this work is to examine the impact of varying $f^{\rm PBH}_{X,\rm esc}$, which is the fraction of X-ray photons escaping the host galaxies, for the lognormal PBH model on the 21-cm signal. This is crucial as this parameter is highly uncertain from both observational and theoretical perspectives. In fact, recent JWST observations of these early AGNs \citep{Maiolino2025_Xray, Yue2024, Ananna_24, Mazzolari_25} estimated a very weak X-ray emission compared to AGNs in the Local Universe. In view of this, we varied $f^{\rm PBH}_{X,\rm esc}$ from 1.0 to 0.2 as shown in Figure-\ref{fig:global_ps_f_X_variation}. The dashed magenta line corresponds to the lognormal model with $f^{\rm PBH}_{X,\rm esc} = 0.2$, while the solid magenta line represents the case with $f^{\rm PBH}_{X,\rm esc} = 1.0$. For comparison, the SF-only scenario is shown by the black curve. We note that the solid magenta and black curves are also presented in Figure-\ref{fig:global_PS}. As $f^{\rm PBH}_{X,\rm esc}$ is lowered from 1.0 to 0.2, the absorption trough becomes progressively deeper, approaching that of the SF-only model. This behaviour is expected, as a lower escape fraction results in reduced X-ray heating of the intergalactic medium, thereby enhancing the depth of the absorption feature. A similar impact can be noticed for the power spectra as in the previous ones. An increasing PBH heating efficiency results in suppression of power spectrum amplitude. Although an efficiency of $f^{\rm PBH}_{X,\rm esc}=0.2$ produces a similar trend as the SF-only case, the effect of a higher efficiency ($f^{\rm PBH}_{X,\rm esc}=1.0$) can be significant enough to be distinguishable by the upcoming interferometers. Since this qualitative trend is similar across all PBH models considered in this work, we restrict our discussion here to the lognormal case.

\section{Conclusions and discussion}
\label{sec:discussion}

In this work, we first applied the PBH analytical model \texttt{PHANES} to three physically motivated PBH mass functions, namely, lognormal, power-law, and critical to compute the key quantities of X-ray heating, Ly-$\alpha$ background, and number of ionising photons, needed to calculate the 21-cm signal. We then extended the explicitly photon-conserving reionisation model \texttt{SCRIPT} to incorporate the essential physics of cosmic dawn. Finally, we coupled the two frameworks to investigate the impact of different PBH mass functions on both the global 21-cm signal and its power spectrum over the redshift range $5 \leq z \leq 30$. Our main findings are as follows:

\begin{itemize}
\item The inclusion of early AGNs leads to a significant impact on both the global 21-cm signal and its power spectrum compared to models that only include SF galaxies. This trend is observed for all PBH mass functions considered in this study.

\item The global signal and the 21-cm power spectrum corresponding to the critical mass function differ significantly from those obtained from lognormal and power-law mass functions.

\item Finally, we examine the effect of varying the escape fraction of X-ray photons from PBH-seeded galaxies in the case of the lognormal mass function. We find that as this fraction decreases, the resulting X-ray heating is reduced, causing the signal to progressively approach the SF-only scenario in the limit of vanishing escape fraction.
\end{itemize}

We conclude this work by noting a few caveats of this study. Most importantly, we note that we have not considered the contribution of radio emission from early AGNs. Such emission could, in principle, lead to a deeper absorption trough by counteracting the effect of enhanced X-ray heating \citep{EwallWice_20, Nelander2025}. However, we did not include this effect, as current observations suggest that these early AGNs produce very weak radio emission, compared to AGNs observed in the Local Universe \citep{Mazzolari_25, Mazzolari_26}. However, we plan to include the radio emission in our future work. Furthermore, we modelled the X-ray heating effects by assuming a simplistic semi-analytic form \citep{furlanetto2006}, which might not be realistic on smaller scales. The model can be improved with more accurate small-scale physics in the future, although the large-scale features are unlikely to change significantly. Furthermore, this framework can also be incorporated with an explicit contribution from Pop-III stars and feedback from Lyman-Werner background \citep{2023MNRAS.520.3609V}, which could play a crucial role in IGM evolution during the cosmic dawn. We note that the current study demonstrates the impact of PBH effects on a fiducial star-formation evolution model, which has significant room for variation, allowing for present observational constraints (i.e. see Appendix \ref{app:f_star_var}, which shows the signal for a couple of other SF variants). We also checked the resolution dependencies of the relevant observables in the semi-numerical setup as a robustness measure (see Appendix-\ref{app:res_comp}). We found that the qualitative conclusions remain the same, even though the different resolutions have slightly different quantitative estimates of the observable effects. As a future work, we are planning to extend this study beyond cosmic dawn to explore the effect even in the context of the dark ages at redshifts beyond $z>30$.

%%%%%%%%%%%%%%%%%%%%%%%%%%%%%%%%%%%%%%%%%%%%%%%%%%%%%%%%%%%%%%
\begin{acknowledgements}
The work of AC was supported by the European Union’s Horizon Europe research and innovation programme under the Marie Skłodowska-Curie Postdoctoral Fellowship HORIZON-MSCA-2023-PF-01, grant agreement No 101151693 (LUPCOS).  The authors acknowledge the use of CHATGPT for refining the text at the final stage of the manuscript.
\end{acknowledgements}

%%%%%%%%%%%%%%%%%%%%%%%%%%%%%%%%%%%%%%%%%%%%%%%%%%%%%%%%%%%%%%
\bibliographystyle{aa} % style aa.bst
\bibliography{main} % your references Yourfile.bib

%%%%%%%%%%%%%%%%%%%%%%%%%%%%%%%%%%%%%%%%%%%%%%%%%%%%%%%%%%%%%%%
% Appendices must be placed after   \end{thebibliography}
% They will be placed automatically on a new page.
%%%%%%%%%%%%%%%%%%%%%%%%%%%%%%%%%%%%%%%%%%%%%%%%%%%%%%%%%%%%%%%
\begin{appendix}
\section{Fiducial reionisation model}
\label{app:reion_hist}
\begin{figure}[ht]
    \sidecaption
    \includegraphics[width=0.5\textwidth]{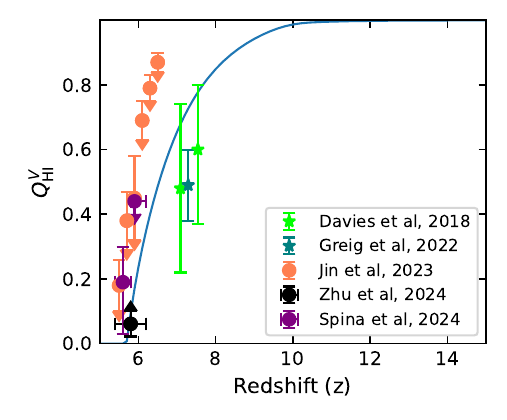}
    \caption{Fiducial ionisation history (a fast evolving case with $f_*\propto (1+z)^{-4}$, obeying existing constraints on the  IGM neutral fraction.}
    \label{fig:ion_hist}
\end{figure}

In Figure-\ref{fig:ion_hist}, we showed the reionisation history of our fiducial model, assumed in this study. We chose a fast reionisation evolution with $f_*=0.005\left(\frac{10}{1+z}\right)^4$, keeping the cosmic dawn window unaffected from photoionisation heating. However, the model obeys different existing constraints on the ionisation histories,  mainly coming from the Ly-$\alpha$ forest \citep{Jin_23, Zhu2024_damping,spina2024} and the damping wing analysis \citep{davies2018, Greig2022} of the high redshift quasar spectra. The model also follows a late reionisation end, which is in accord with the recently emerging scenario \citep{bosman2022}.

\section{Effect of varying $f_{*,0}$ on the 21-cm signal}
\label{app:f_star_var}
\begin{figure}[ht]
    \sidecaption
    \includegraphics[width=\linewidth]{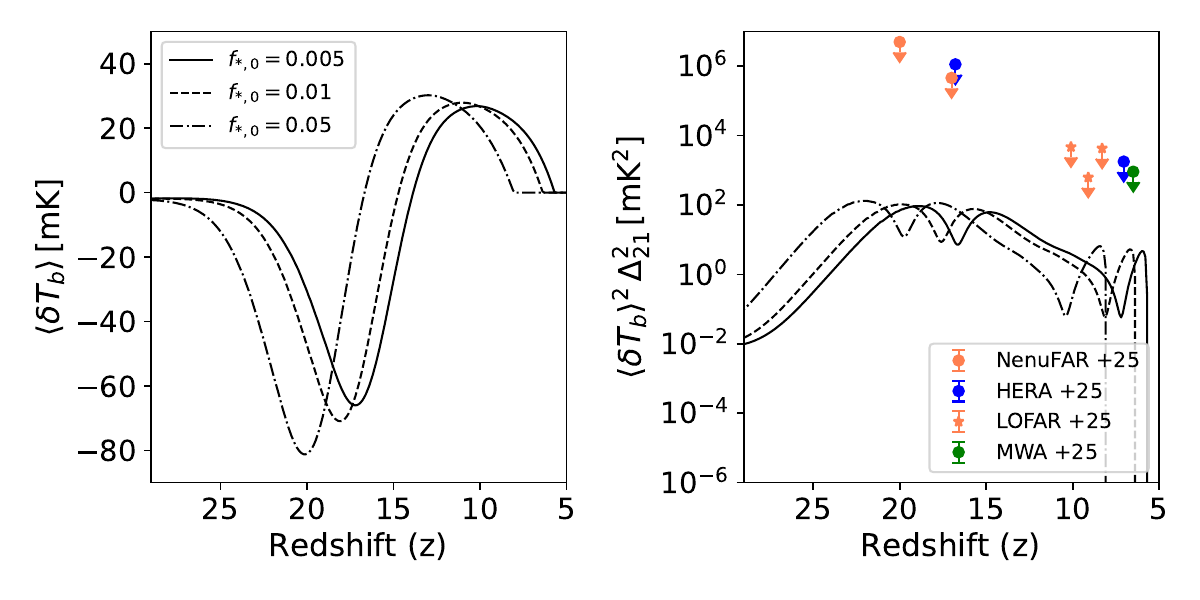}
    \caption{Redshift evolution of the Global 21-cm signal and the corresponding power spectra ($k\sim0.1~{h/\rm cMpc}$) for different $f_{*, 0}$, keeping $f^{\rm SF}_{X, \rm esc}$ fixed at $1.0$.}
    \label{fig:fstar}
\end{figure}

We have shown in Figure-\ref{fig:fstar}, 21-cm signal (the global as well as the power spectra) with varying the $f_{*,0}$. This exercise has been performed to check whether the newly extended \texttt{SCRIPT} framework is implemented correctly. The left panel shows the global signal. As can be seen, with higher value of $f_{*,0}$, the absorption trough appears at earlier redshift. For example, with $f_{*,0}=0.05$, the trough appears at $z \sim 20$, whereas for $f_{*,0}=0.005$, the absorption trough appears at $z \sim 18$. Further models with increasing $f_{*,0}$ also reaches zero at earlier redshift since reionisation is completed earlier in those models. Similar effect has also been observed for the power spectrum, shown in the right panel of Figure-\ref{fig:fstar}. Here, all the peaks (related to Ly-$\alpha$, X-ray heating and reionisation), has shifted to earlier redshift as we increase the value of $f_{*,0}$.

\section{Comparison between resolutions}\label{app:res_comp}
\begin{figure}[ht]
    \sidecaption
    \includegraphics[width=\linewidth]{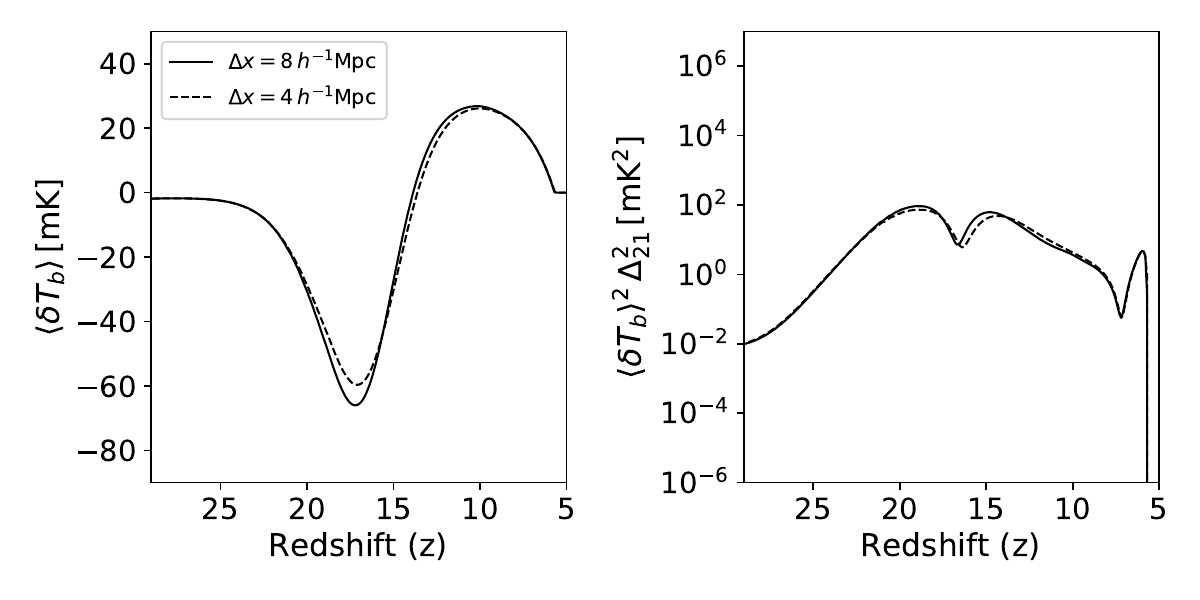}
    \caption{Comparison of 21-cm observables between two resolutions ($\Delta x=4 ~h^{-1}\rm cMpc$ and $8 ~h^{-1}\rm cMpc$) of the simulation setup.}
    \label{fig:ion_hist_resolution}
\end{figure}

 In Figure-\ref{fig:ion_hist_resolution}, we show the comparison of the global 21-cm signal along with power spectra, for two different resolutions of the box (i.e. fiducial with $\Delta x =8~h^{-1}\rm cMpc$ and a higher resolution variant with $\Delta x =4~h^{-1}\rm cMpc$).  We note that the differences are maximum near the trough and agree well with each other for most of the redshifts. In both the cases, we assumed $f_{*,0}=0.005$ and $f^{\rm SF}_{X, \rm esc} =1.0$.

\end{appendix}

\label{lastpage}
\end{document}